\documentclass[12pt]{article}
\usepackage{latexsym,epsfig,graphicx,a4wide,amssymb}

{\rm }

\def\be{\begin{equation}}
 \def\ee{\end{equation}}
 \def\bea{\begin{eqnarray}}
 \def\eea{\end{eqnarray}}

\def\2{\frac{1}{2}}
\def\4{\frac{1}{4}}

\catcode`\@=11

\@addtoreset{equation}{section}

\def\@normalsize{\@setsize\normalsize{15pt}\xiipt\@xiipt
\abovedisplayskip 14pt plus3pt minus3pt%
\belowdisplayskip \abovedisplayskip
\abovedisplayshortskip  \z@ plus3pt%
\belowdisplayshortskip  7pt plus3.5pt minus0pt}
\def\small{\@setsize\small{13.6pt}\xipt\@xipt
\abovedisplayskip 13pt plus3pt minus3pt%
\belowdisplayskip \abovedisplayskip
\abovedisplayshortskip  \z@ plus3pt%
\belowdisplayshortskip  7pt plus3.5pt minus0pt
\def\@listi{\parsep 4.5pt plus 2pt minus 1pt
            \itemsep \parsep
            \topsep 9pt plus 3pt minus 3pt}}

\def\underline#1{\relax\ifmmode\@@underline#1\else
        $\@@underline{\hbox{#1}}$\relax\fi}
\@twosidetrue \relax

\catcode`@=12

\catcode`\@=11

\def\section{\@startsection{section}{1}{\z@}{3.5ex plus 1ex minus
   .2ex}{2.3ex plus .2ex}{\large\bf}}

\def\ps@headings{\def\@oddfoot{}\def\@evenfoot{}
\def\@oddhead{\hbox{}\hfill
        \makebox[.5\textwidth]{\raggedright\ignorespaces --\thepage{}--
        \hfill }}
\def\@evenhead{\@oddhead}
\def\subsectionmark##1{\markboth{##1}{}}
}

\ps@headings

\catcode`\@=12

\begin{document}

\begin{titlepage}
%
%


%

\begin{centering}
\vspace{1cm}
{\Large {\bf Black hole solutions in Horava-Lifshitz Gravity with cubic terms}}\\

\vspace{1.5cm}

{\bf George Koutsoumbas $^{\sharp}$}, {\bf Pavlos Pasipoularides
$^{*}$} \\
 \vspace{.2in}
 Department of Physics, National Technical University of
Athens, \\
Zografou Campus GR 157 73, Athens, Greece \\

\vspace{3mm}

\end{centering}
\vspace{1.5cm}

\begin{abstract}
We study four dimensional non-projectable Horava-Lifshitz type gravity, in the case of an
action with terms, cubic in curvature. For special choices of the free
parameters of the model, we obtain two new analytic black hole solutions which exhibit the standard Schwarzschild
asymptotic behavior in the large distance limit. The effect of cubic terms in the short range
 behavior of the black hole solutions is discussed.
\end{abstract}

\vspace{1.5cm}
\begin{flushleft}

$^{\sharp}~~$ e-mail address: kutsubas@central.ntua.gr \\ $^{*}
~~$ e-mail address: paul@central.ntua.gr \\

\end{flushleft}
\end{titlepage}

\section{Introduction}

A new higher order gravity model, which claims power-counting
renormalizability and it serves as a UV completion of general relativity,
has been formulated recently by Horava~\cite{Horava:2008ih}. This scenario is based on an anisotropy
between space and time coordinates, which is expressed via the
scalings $t\rightarrow b^z t$ and  $x\rightarrow b x$, where $z$
is a dynamical critical exponent. Note, that in the Horava-Lifshitz (HL) gravity the
four-dimensional diffeomorphism  invariance of general relativity
is sacrificed in order to achieve power-counting
renormalizability.

It is worth noting that HL gravity has stimulated an extended research
on cosmology and black hole solutions
\cite{Kiritsis:2009sh}-\cite{Eune:2010kx}.
In addition to general relativity studies, quantum field theory
models in flat anisotropic space-time have also been
considered \cite{Visser:2009fg}-\cite{Romero:2010tc}.

As it has already been mentioned, the HL gravity violates local
Lorentz invariance in the UV, however it is expected that general
relativity is recovered in the IR limit. This implies a very
special renormalization group flow for the couplings of the model,
but there is no theoretical study supporting this specific
behavior. In addition, there are several other possible inconsistencies in
HL gravity which have been discussed (see for example
\cite{Charmousis:2009tc}-\cite{Pons:2010ke}
and references therein). In particular, the absence of full diffeomorphism invariance
introduces an additional scalar mode which may lead to strong
coupling problems or instabilities, and in this way it prevents HL gravity to fully reproduce general relativity in the IR limit.
However, these problems will not be addressed in the present paper.

For the construction of the HL action, there has been
proposed \cite{Horava:2008ih}
the so called "detailed balance principle".
The main advantage of this approach is the restriction of the
large number of arbitrary couplings that
appear in the action of the model. However,
a more general way for constructing the action
would be to include all possible operators which are compatible
with the renormalizability \cite{Kiritsis:2009sh,Sotiriou:2009bx}; this implies
that all operators with dimension less or equal to six are allowed.

Before proceeding it is important to mention that the HL gravity
can be separated into two versions which are known as projectable
and non-projectable. In the projectable version the lapse function
$N$ (see Eq. (\ref{ADM}) below) depends only on the time coordinate, while in the
non--projectable version $N$ is a function of both the space and
time coordinates.

In this paper we study spherically symmetric solutions for non-projectable Horava-Lifshitz gravity, in the case of an
action without detailed balance, that includes all possible terms which are quadratic and cubic in curvature.
For special choices of the free parameters of the model, we obtain two analytical solutions which
exhibit the standard Schwarzschild black hole asymptotic behavior in the large distance limit.
The effect of the cubic terms in the short range behavior of the black hole solutions is discussed.

One should refer to some previous works towards this direction. In the case
of actions with detailed balance, spherically symmetric solutions were
found \cite{Lu:2009em}, but they had an unconventional large
distance asymptotic behaviour. The correct Schwarzchild-flat
asymptotic behaviour can be recovered if the detailed balance
action is modified in the IR by a term proportional to the Ricci
scalar, and the cosmological constant term is set to
zero \cite{Kehagias:2009is}. A similar study,  in the case of a
non-vanishing cosmological constant, has also been carried out
\cite{Park:2009zra}. A generalization to topological black holes
has been obtained in  \cite{Cai:2009pe}. In addition, a systematic study of
static spherically symmetric solutions of HL gravity (for an
action which includes only the quadratic terms in curvature) has been
presented in \cite{Kiritsis:2009rx} where the most general
spherically symmetric solution for general coupling parameters was obtained.
Finally, a general study of black hole solutions in the case of 5D Horava-Lifshitz Gravity,
with quadratic terms in curvature, is presented in Ref. \cite{Koutsoumbas:2010pt}.
Note that in this work we consider for the first time the issue of
spherically symmetric solutions for an action including terms cubic in curvature.

The paper is organized as follows. In Sec. 2 we write down the most
general action for HL gravity. In Sec. 3 we derive the equations of
motion. In Sec. 4 we present the two static spherically symmetric
solutions and finally Sec. 5 contains our conclusions. In Appendix \ref{appA}
we show that the above mentioned two solutions, when the cubic terms are absent, coincide
with solutions which are already known in the literature. Finally, in Appendix \ref{appB} we give a
brief presentation of detail balance condition, and we discuss it in connection with the results of Appendix \ref{appA1}.

\section{Horava-Lifshitz Gravity}

In this section we introduce the notation for the so-called Horava
gravity models in the case of three spatial dimensions ($d=3$).
These models are characterized by an anisotropy between space and
time dimensions
\begin{equation}
[t]=-z, \quad [x]=-1~,
\end{equation}
where $z$ is an integer dynamical exponent. In order to derive the
action of the model, it is useful to express the space-time metric
in the Arnowitt, Deser and Misner (ADM) form
\begin{equation} \label{ADM}
ds^2=-  c^2 N^2 dt^2+g_{ij} \left(dx^i-N^i dt\right)\left(dx^j-N^j
dt\right)~,
\end{equation}
where $c$ is the
velocity of light, with dimension $[c]=z-1$, and spatial
components $dx^i/dt$ $(i=1,2,3)$. In addition, $N$ and $N_i$ are
the "lapse" and "shift" functions which are used in general
relativity in order to split the space-time dimensions, and
$g_{ij}$ is the spatial metric of signature (+,+,+).  For the
dimensions of "lapse" and "shift" functions we obtain
\begin{equation}
[N]=0, ~~[N_i]=z-1~.
\end{equation}  The action of the model is constructed from a
kinetic plus a potential term according to the equation \be
\label{action}
 S=\frac{1}{16 \pi G c }\int dt d^d x \sqrt{|g|} N \left \{ {\cal L}_{K}+{\cal L}_{V}\right\}
\ee in which $d$ ($D=d+1=4$) is the spatial dimension and $G$ is
the Newton constant.

In this paper the dynamical exponent $z$ is set
equal to $3$. Note, that this choice for $z$ is an immediate consequence of
power counting renormalizibility request. In particular, the coupling $16 \pi G c$ in the above action
has dimension $[16 \pi G c]=z-3$, hence, if $z=3$ the HL gravity model is renormalizable, for $z>3$ is superrenormalizable and
for $0<z<3$ is nonrenormalizable.

We emphasize that the main motivation for considering models of this
type is the construction of a power-counting renormalizable
gravity model. However,  in order to achieve renormalizibility,
and simultaneously keep the time derivatives up to second order,
we have to sacrifice the standard 4D diffeomorphism invariance of
general relativity, which is now restricted to the transformation
\be \label{diffeo} \tilde{x}^{i}=\tilde{x}(x^j,t), ~~
\tilde{t}=\tilde{t}(t)~. \ee The kinetic part in the above
Lagrangian of Eq. (\ref{action}) can be expressed via the
extrinsic curvature as: \be \label{curvature1}
 {\cal L}_{K}= (K^{ij}K_{ij}-\lambda K^2), \quad K_{ij}=\frac{1}{2 N} \left\{-\partial_{t}g_{ij}+\nabla_i N_j+\nabla_j N_i\right\},~~ i,j=1,2,3
\ee which is invariant under the transformations of Eq.
(\ref{diffeo}). Note, that $\lambda$ is a dimensionless running coupling which is expected to
approach unity in the IR limit, so local Lorentz invariance is restored, as we explain later in detail.
For the construction of the potential term we will
not follow the standard detailed balance principle (for details see Sec. 7), but we will
use the more general approach \cite{Kiritsis:2009sh, Sotiriou:2009bx}, according to
which the potential term is constructed by including all possible
renormalizable operators \footnote{We have ignored the terms which
violate parity, see also \cite{Sotiriou:2009bx}.}, that have dimension smaller
than or equal to six, hence we write
\be
 {\cal L}_{V}= {\cal L}_R+{\cal L}_{R^2}+{\cal L}_{R^3}+{\cal L}_{\Delta R^2}
\ee
where
\bea
&& {\cal L}_R=\eta_{0a}+\eta_{1a} R, \quad {\cal L}_{R^2}=\eta_{2a} R^2+\eta_{2b} R^{ij}R_{ij} \\
&& {\cal L}_{R^3}=\eta_{3a} R^3+\eta_{3b} R R^{ij} R_{ij}+\eta_{3c} R_i^j R_j^k R^i_k\\
&&{\cal L}_{\Delta R^2}=\eta_{3d} R\nabla^{2}R+ \eta_{3e}\nabla_i R_{jk} \nabla^i R^{jk}
\eea
The dimensions of the various terms in the Lagrangian read
\be [R]=2,~ [{R^2}]=4, ~[{R^3}]=[{\Delta R^2}]=6. \ee
where the symbol $\Delta$ is defined as $\Delta=g_{ij}\partial_i\partial_j$ ($i=1,2,3$).
In addition, we have used the notation $R$, $R_{ij}$ and $R_{ijkl}$
for the Ricci scalar,  the Ricci and the Riemann tensors
($i,j=1,2,3$), which correspond to the spatial 3D metric
$g_{ij}$. Note that the term $ R^{ijkl}R_{ijkl}$ does not appear in ${\cal L}_{R^2}$, as the Weyl tensor in three
dimensions automatically vanishes.

The first term ${\cal L}_R$, with $\eta_{0a}=0$, is necessary in order to recover
general relativity, without a cosmological constant term \footnote{We are interested for spherically symmetric solutions which have the standard Schwarzschild asymptotic behavior, so we have set the cosmological constant $\eta_{0a}$ equal to zero}, in the IR limit.
The other terms ${\cal L}_{R^2}$, ${\cal L}_{R^3}$, ${\cal L}_{\Delta R^2}$ include all possible quadratic and cubic
terms in curvature, which become important in short distances when the space-time curvature becomes large. In addition, $\eta_{2a}$, $\eta_{2b}$,
$\eta_{3a}$, $\eta_{3b}$, $\eta_{3c}$, $\eta_{3d}$, $\eta_{3e}$ are coupling constants
with dimensions
\be [\eta_{1a}]=6, ~~
[\eta_{2a}]=[\eta_{2b}]=4,~~
[\eta_{3a}]=[\eta_{3b}]=[\eta_{3c}]=[\eta_{3d}]=[\eta_{3e}]=0.\ee

If this model is to make sense, it is necessary that general relativity is
recovered in the IR limit. Although
this difficult question has not been analyzed, we will assume that the
renormalization group flow towards the IR leads the parameter
$\lambda$ very close to the value one ($\lambda=1$), hence general
relativity is recovered \footnote{As we have note in the introduction, a scalar mode remains, in the IR limit, due to the absent of local Lorentz invariance. This mode
is responsible for instabilities and strong coupling problems.}. Also, to obtain the Einstein-Hilbert
action \be \label{IR}
 S_{EH}= \frac{1}{16\pi G} \int dx^0 d^3 x \sqrt{g} N
\left (\tilde{K}_{ij}\tilde{K}^{ij}- \tilde{K}^2 + R
\right)~, \ee without a cosmological constant term, we have to set $\eta_{1a}=c^2$, $\eta_{0a}=0$, and
\be
\tilde{K}_{ij}=\frac{1}{2 N} \left\{-\partial_{0}g_{ij}+\nabla_i
\left(\frac{ N_j}{c}\right)+\nabla_j\left(\frac{
N_i}{c}\right)\right\}~. \ee where the time-like coordinate
$x_{0}$ is defined as $x_{0}=c t$.

\section{Equation of motion}

We are looking for spherically symmetric solution, in the case of non-projectable
Horava-Lifshitz type gravity when the lagrangian part
$${\cal L}_{\Delta R^2}=\eta_{3d} R\nabla^{2}R+ \eta_{3e}\nabla_i R_{jk} \nabla^i R^{jk}$$ vanishes,
or equivalently $\eta_{3d}=\eta_{3e}=0$.
As we will see, this assumption simplifies significantly the corresponding equations of motion.
For the metric \footnote{There is a more general
ansatz for the metric, of the form $ds^2=-N(r)^2
dt^2+f^{-1}(r)~(dr+N^{r}(r)dt)^2 +r^2 d\Omega^2$ with nonzero
shift $N^{r}(r)$ (see Ref. \cite{Capasso:2009ks}), but we have set $N^{r}(r)=0$
( see Ref. \cite{Kiritsis:2009rx}) to simplify the equations.} we make the ansatz:
\be \label{metric}
 ds^2=-N(r)^2 dt^2+\frac{dr^2}{f(r)} +r^2 d\Omega^2
\ee
where $r$ is the radius coordinate and
\be
 d\Omega^2=d\theta^2+\sin^2\theta d \phi^2
\ee
is the metric of a 2D maximally symmetric space,with spherical topology.
In what follows it is convenient to perform the transformation
\be
f(r)=1+r^2 Z(r) \label{Z}.
\ee
The action of the model, after angular integration, can be put in the form
\be
S\left[N(r),Z(r),\frac{dZ(r)}{dr}\right]=\int_{0}^{+\infty} dr~  L\left[N(r),Z(r),\frac{dZ(r)}{dr}\right]
\ee
where
\be
L\left[N,Z,\frac{dZ}{dr}\right]\sim r^2\sqrt{\frac{N^2}{f}}
\left(R\left(r \frac{d Z}{dr}\right)^3+P(Z)\left(r \frac{d Z}{dr}\right)^2+ 4 M(Z) \left(r \frac{d Z}{dr}\right)+ 4 Q(Z)\right)
\ee
and the coefficients $R$, $M(Z),P(Z)$ and $Q(Z)$ are determined by the equations
$$R=-\frac{5}{2}(32 \eta_{3a}+12 \eta_{3b}+5\eta_{3c})$$ $$M(Z)=-(540 \eta_{3a}+180 \eta_{3b}+60 \eta_{3c}) Z^2+(60 \eta_{2a}+20 \eta_{2b})Z-5 \eta_{1a} ,$$ $$P(Z)= -(720 \eta_{3a} + 250 \eta_{3b}+90 \eta_{3c})Z+(40 \eta_{2a}+15 \eta_{2b})$$
$$Q(Z)=-(540 \eta_{3a}+180 \eta_{3b}+60 \eta_{3c}) Z^3+(90 \eta_{2a} + 30 \eta_{2b})Z^2-15 Z \eta_{1a}$$
Note, that the coefficient $R$ is a constant independent from $r$, while the functions $M(Z),P(Z)$ and $Q(Z)$ do not depend explicitly on $r$.
In the case of the spherically symmetric ansatz, it is convenient to define five new couplings according to the relations
\bea
&&\rho=32 \eta_{3a}+12 \eta_{3b}+5\eta_{3c}, ~~\eta=540 \eta_{3a}+180 \eta_{3b}+60 \eta_{3c} \nonumber\\ && \zeta=720 \eta_{3a} + 250 \eta_{3b}+90 \eta_{3c},~~\tau=30 \eta_{2a}+10 \eta_{2b},~~\varphi=40 \eta_{2a}+15 \eta_{2b},
\eea
so we obtain
\bea && R=-\frac{5}{2}\varrho \nonumber \\
     &&M(Z)=-\eta Z^2+2 \tau Z-5 \nonumber \\
     &&P(Z)= -\zeta Z+\varphi \nonumber \\
     &&Q(Z)=-\eta Z^3+3 \tau Z^2-15 Z
\eea
The corresponding equations of motion read:
\bea
\frac{d}{dr}\left(\frac{\partial L}{\partial N'}\right) -\frac{\partial L}{\partial N}=0, ~~~ \frac{d}{dr}\left(\frac{\partial L}{\partial Z'}\right) -\frac{\partial L}{\partial Z}=0
\eea
The first of the Euler-Lagrange equations yields for the function $Z(r)$ the equation
\be
R\left(r \frac{d Z}{dr}\right)^3+P(Z)\left(r \frac{d Z}{dr}\right)^2+ 4 M(Z) \left(r \frac{d Z}{dr}\right)+ 4 Q(Z)=0 \label{Zeq}
 \ee
If we solve the above equation we obtain a first order differential equation
\be
r \frac{d Z}{dr}=H_{\ell}(Z), ~~~\ell=1,2,3
\ee
In general the algebraic equation (\ref{Zeq}) has three solutions $H_{\ell}(Z),(\ell=1,2,3)$,
however a function $H_{\ell}(Z)$ is acceptable only if the following
three conditions  are satisfied
 \begin{itemize}
   \item $H_{\ell}(Z(r))$ should be a real function for all the values of $r$ in the range: $0<r<+\infty$
   \item $H_{\ell}(Z)\simeq -3Z+O(Z^2)$ for $Z\rightarrow 0$, this implies that $f(r)$ behaves as $f(r)\simeq 1-\frac{2M}{r}$ asymptotically
   \item $N(r)^2\simeq f(r) $ for $r\rightarrow +\infty$, if we demand the metric to recover the standard Schwarzschild form asymptotically
 \end{itemize}
Notice that there are solutions that violate the above three condition. For example, there are solutions with Anti de Sitter (AdS) or de Sitter (dS) asymptotic behavior.
However they will not be examined in the present paper. We will now derive the equation of motion for the function $N(r)$. If we set
\be
 \bar{N}(r)^2=\frac{N(r)^2}{f(r)},
\ee
from the second of the Euler-Lagrange equations above we obtain
\be \label{N1}
\frac{d \bar{N}(r)}{d r} + \bar{C}(r) \bar{N}(r)=0,
~~~ \bar{C}(r) =\left[\frac{1}{r^3 G_1} \frac{d (r^3 G_1)}{d r}-\frac{G_2}{rG_1}\right] \label{Neq} \ee
where
\bea  && G_1 = 3 R \left(r\frac{d Z}{d r}\right)^2+ 2
 P(Z) \left(r\frac{d Z}{d r}\right) + 4  M(Z), \\ &&G_2= P'(Z) \left(r\frac{d Z}{d r}\right)^2 + 4  M'(Z)\left( r\frac{d Z}{d
r}\right) + 4 Q'(Z). \eea
It is convenient to change the variable in Eq. (\ref{Neq}) from $r$ to $Z$, to obtain:
\be \label{g01}
\frac{d \tilde{N}(Z)}{d Z} + \tilde{C}(Z) \tilde{N}(Z)=0,
~~~ \tilde{C}(Z) =\frac{1}{H(Z)}\left[3 -\frac{\tilde{G}_2}{\tilde{G}_1}\right]+\frac{1}{\tilde{G}_1}\frac{d  \tilde{G}_1}{dZ} \label{NeqZ} \ee
where
\bea \label{gZ1} && \tilde{G}_1(Z) = 3 R H(Z)^2+ 2
 P(Z) H(Z) + 4  M(Z), \\\label{gZ2}  &&\tilde{G}_2(Z)= P'(Z)H(Z)^2 + 4  M'(Z)H(Z) + 4 Q'(Z). \eea
Finally, we emphasize that the parameter $\lambda$ does not appear
in the equations of motion, so the spherically symmetric solutions depend only on the
parameters $\varrho,~ \eta,~ \tau, ~\zeta$ and $\varphi$. The reason is that we are looking for
static spherically symmetric solutions solutions with zero shift functions, hence extrinsic curvature terms do not
contribute to the equations of motion, as they contain only time
derivatives of the metric components (in particular $K_{ij}=0$ for the metric  of Eq. (\ref{metric})). In addition, the
parameter $\lambda$ appears only in the extrinsic curvature part
of the action. Therefore, the solutions we obtain in the
following sections will be valid for arbitrary values of
$\lambda$.

\section{Analytic solutions}

Although, it is not possible to give a full analytical treatment of the most general case in the presence of terms both quadratic as well as cubic in curvature, we have obtained analytic solutions for two
specific choices of the couplings of the model, which are presented in detail in the following subsections \ref{sec41} and \ref{sec42}.

\subsection{First case: $\varrho=0$, $\zeta=0$ and $\varphi=0$}\label{sec41}

An analytical treatment of a third order algebraic equation (see Eq. (\ref{Zeq})) is a difficult task.
However, if we set $\rho=\zeta=\varphi=0$, we can reduced this equation to a first order equation which can be handle analytically. In this case we have
$R=0$ and $P(Z)=0$, hence Eq. (\ref{Zeq}) can put into the form
\be
r\frac{dZ}{dr}=-\frac{Q(Z)}{M(Z)} \label{Zeqfirst}
\ee
where $Q(Z)$ and $M(Z)$ are given by the equations
\begin{eqnarray}\label{MQ}
&&M(Z)=-\eta Z^2+2 \tau Z-5\\
&&Q(Z)=-\eta Z^3+3 \tau Z^2-15Z
\end{eqnarray}
Now, if we integrate Eq. (\ref{Zeqfirst}) we find
\be
Z\left(\eta Z^2-3 \tau Z+15\right)=-\frac{\tilde{C}_{M}}{r^3} \label{Zeq0}
\ee
where $\tilde{C}_{M}$ is an integration constant which is related with the mass of the black hole, as we will see later.

For the general case in which $\eta \neq 0$, we have to solve the following third order algebraic equation
\be
\eta Z^3-3 \tau Z^2+15 Z+\frac{\tilde{C}_{M}}{r^3}=0 \label{cubicZ}
\ee
If we set
\be \label{defqw}
q=\frac{5\eta-\tau^2}{\eta^2}, ~~w=\frac{2 \tau^3-15 \tau \eta}{2 \eta^3}-\frac{\tilde{C}_M}{2 \eta r^3},
\ee
the discriminant $\Delta_{3}=q^3+w^2$, of the cubic Eq. (\ref{cubicZ}), is
\be
\Delta_3=\left(\frac{5\eta-\tau^2}{\eta^2}\right)^3+\left(\frac{2 \tau^3-15 \tau \eta}{2 \eta^3}-\frac{\tilde{C}_M}{2 \eta r^3}\right)^2
\ee
After some algebraic computation we find
\be \label{disc3}
\Delta_3=\frac{25\left(20\eta-3 \tau^2\right) X^2+2 \tau  \tilde{C}_{M} (15 \eta-2 \tau^2) X+\eta^2 \tilde{C}_{M}^2}{4 \eta^2 X^2}
\ee
where we have set $X=r^3$. The numerator of $\Delta_3$ is a quadratic polynomial with discriminant
\be \label{disc2}
\Delta_2=-16 \tilde{C}_{M}^2(5\eta-\tau^2)^3
\ee
According to the behavior of $\Delta_3$ we have three distinct cases:
\begin{enumerate}
  \item Non negative determinant $\Delta_3$ in the range $0<r<+\infty$, equivalently $\Delta_3=0$ has no real root for $X=r^3$ or it has two negative real roots  ($\tau^2 \leq 5 \eta$, or $5 \eta<\tau^2< \frac{20}{3} \eta$ and $\tau \tilde{C}_{M} >0$).
  \item One real positive and one real negative root ($\frac{20}{3}\eta<\tau^2$).
  \item Two real positive roots: ($5 \eta<\tau^2< \frac{20}{3} \eta$ and $\tau \tilde{C}_{M}<0$).
\end{enumerate}
which are analyzed in the following three subsections. Note that although the construction of spherically symmetric solutions in these three cases appears
different, the characteristic features of the solutions are similar, as discussed in subsection 4.1.8.

\subsubsection{Non negative discriminant $\Delta_3$ in the range $0<r<+\infty$}

In this section we will study the case for which $\Delta_3 \geq 0$ in the range  $0<r<+\infty$. The discriminant $\Delta_3$ (see Eq. (\ref{disc3})) is non-negative when $\Delta_2\leq 0$ (see Eq. (\ref{disc2})) and the coefficient $20\eta-3 \tau^2>0$, of $X^2$ in Eq. (\ref{disc3}), is positive. These conditions are equivalent to $5 \eta-\tau^2\geq 0$. An alternative way to achieve non-negative $\Delta_3$ for $X=r^3>0$ is the case of two negative roots, which corresponds to the condition $5 \eta<\tau^2< \frac{20}{3} \eta$ and $\tau>0$. Hence, the solutions we present in this subsection are valid only when the parameters $\eta$ and $\tau$ obey the restrictions $\tau^2 \leq 5 \eta$, or $5 \eta<\tau^2< \frac{20}{3} \eta$ and $\tau \tilde{C}_{M}>0$.

In the case of non-negative discriminant $\Delta_3$, as we know from the theory of cubic algebraic equations, there is a unique real solution \footnote{If the discriminant is positive there is one real and two complex conjugate solutions. If the discriminant is zero we have one triple real solution.} which can be expressed as
\begin{equation} \label{sol1}
Z(r)=s+t+\frac{\tau}{\eta}
\end{equation}
where
\be
s=\sqrt[3]{w+\sqrt{\Delta_3}},~~t=\sqrt[3]{w-\sqrt{\Delta_3}}
\ee
or equivalently
\be \label{sval} s=\left[\frac{2 \tau^3-15 \tau \eta}{2 \eta^3}-\frac{\tilde{C}_M}{2 \eta r^3}
+\sqrt{\left(\frac{5\eta-\tau^2}{\eta^2}\right)^3+\left(\frac{2 \tau^3-15 \tau \eta}{2 \eta^3}-\frac{\tilde{C}_M}{2 \eta r^3}\right)^2}\right]^{1/3}\ee
\be \label{tval} t=\left[\frac{2 \tau^3-15 \tau \eta}{2 \eta^3}-\frac{\tilde{C}_M}{2 \eta r^3}
-\sqrt{\left(\frac{5\eta-\tau^2}{\eta^2}\right)^3+\left(\frac{2 \tau^3-15 \tau \eta}{2 \eta^3}-\frac{\tilde{C}_M}{2 \eta r^3}\right)^2}\right]^{1/3}\ee
while for the function $f(r)$ we take
\be
f(r)=1+ r^2  \left(s+t+\frac{\tau}{\eta}\right)
\ee

\subsubsection{$\frac{20}{3}\eta<\tau^2 $}

For $\frac{20}{3}\eta<\tau^2$ the discriminant $\Delta_{3}$ has one positive and one negative root. Note that only
in this case the coupling $\eta$ can be negative. If $r_{1}$ is the positive root of $\Delta_{3}$, we find that $\Delta_{3}<0$ for $r>r_{1}$ and $\Delta_{3}\geq 0$ for $0< r\leq r_{1}$.
We remind the reader that for negative $\Delta_{3}$ (for  $r>r_{1}$) the cubic algebraic equation (\ref{cubicZ}) has three real solutions. In particular we have:
\begin{eqnarray}
&&Z_1=\tilde{s}+\tilde{t}+\frac{\tau}{\eta}\\
&&Z_2=-\frac{1}{2}(\tilde{s}+\tilde{t})+\frac{\sqrt{3}}{2}(\tilde{s}-\tilde{t})i+\frac{\tau}{\eta}\\ \label{solz2}
&&Z_3=-\frac{1}{2}(\tilde{s}+\tilde{t})-\frac{\sqrt{3}}{2}(\tilde{s}-\tilde{t})i+\frac{\tau}{\eta}
\end{eqnarray}
with $\tilde{s}$ and $\tilde{t}$ defined through
\begin{equation}
\tilde{s}=\sqrt[3]{\rho} e^{i\frac{\theta}{3}},~~\tilde{t}=\sqrt[3]{\rho} e^{-i\frac{\theta}{3}}
\end{equation}
where $\rho$ and $\theta$ are given by the equations
\begin{equation}
\rho=\sqrt{-q^3}, ~~\theta=\cos^{-1}\left(\frac{w}{\rho}\right)
\end{equation}
For the definitions of $q$ and $w$, see Eq. (\ref{defqw}) above. Note
that $q^2<0$, since we examine the case of negative discriminant $\Delta_3=q^3+w^2$.

We have checked that only the second solution $Z_2$ leads to the standard Schwarzschild asymptotic behavior, while the
other two $Z_1$ and $Z_3$ have an $AdS$ or $dS$ large $r$ asymptotic behavior. So, for the construction of spherically symmetric solutions of the form $f(r)=1+r^2 Z$,
we will use the function $Z_2$ for $r>r_{1}$, and the function $Z$, Eq. (\ref{sol1}) of the previous section, for $r\leq r_{1}$. Also, we would like to stress that the
function $f(r)$ is continuous and has a continuous first derivative on the matching point $r_{1}$, as required. On the other hand, if had used
$Z_1$ or $Z_3,$ rather than $Z_2,$ the function $f'(r)$ would have been discontinuous at $r_{1}$.

\subsubsection{$5 \eta<\tau^2< \frac{20}{3} \eta$ and $\tau \tilde{C}_{M}<0$}

In this case the discriminant $\Delta_{3}$ has two positive roots, which are
denoted as $r_1$ and $r_2$ ($r_1<r_2$). If $r$ is inside the
interval $(r_1,r_2)$ the discriminant $\Delta_{3}$ is negative,
while for r outside this interval the discriminant is positive.
Similarly with the previous section
we can construct spherically symmetric solutions, which are
continuous with continuous first derivatives at the matching
points $r_1$ and $r_2$, by assuming that $Z(r)$ is given by
Eq. (\ref{solz2}), if $r_1 \leq r\leq r_2$, and by Eq. (\ref{sol1}),
if $0<r<r_1$ or $r>r_2$.

\subsubsection{Determination of $N(r)^2$}

As discussed in the previous section, the Euler-Lagrange equations for $N$ yield
\be
\frac{d \tilde{N}(Z)}{d Z} + \tilde{C}(Z) \tilde{N}(Z)=0,
~~~ \tilde{C}(Z) =\frac{1}{H(Z)}\left[3 -\frac{\tilde{G}_2}{\tilde{G}_1}\right]+\frac{1}{\tilde{G}_1}\frac{d  \tilde{G}_1}{dZ} \label{NeqZ1} \ee
where $$H(Z)=-\frac{Q(Z)}{M(Z)},$$
$$\tilde{G}_1 =  4 M(Z),$$ $$ \tilde{G}_2 = 4 \frac{Q(Z) M'(Z) }{M(Z)} + 4 Q'(Z).$$
Then we obtain
$$\tilde{C}(Z)=\frac{Q'(Z)-3M(Z)}{Q(Z)}=0,$$
where we have taken into account equations (\ref{MQ}) for $M(Z)$ and $Q(Z)$.
Finally, with an appropriate choice of the integration constant, we find
\begin{equation}
N(r)^2=f(r)
\end{equation}

\subsubsection{Large distance limit}

For $r\rightarrow +\infty$ the asymptotic behavior of the function $f(r)$ should be identified with
the one of the Schwarzschild solution:
\be \label{asymscharz}
r\rightarrow +\infty \Rightarrow f(r)\simeq 1-\frac{2 M }{r}, ~~
\ee
where the Newton constant $G_{N}$ has been absorbed in the mass $M$ ($M\leftrightarrow M G_{N}$) .
If we compare with Eq. (\ref{Z}) we obtain the corresponding asymptotic behavior for the function $Z(r)$ when $r\rightarrow +\infty$
\be\label{asymscharzZ}
r\rightarrow +\infty \Rightarrow  Z(r) \simeq -\frac{2 M }{r^3}
\ee
In order to find the asymptotic behavior of $f(r)$ we can use Eq. (\ref{cubicZ}). For $Z\rightarrow 0$, if we neglect the higher order terms $Z^2$ and $Z^3$ in Eq. (\ref{cubicZ}), we obtain:
\be
Z(r) \simeq -\frac{\tilde{C}_{ M} }{15 r^3}+O\left(\frac{1}{r^6}\right)
\ee
Now, if we compare with Eq. (\ref{asymscharzZ}) we find that $2 M=\tilde{C}_{ M}/15$.

\subsubsection{Short distance limit}

Note that for $r\rightarrow 0$ the discriminant $\Delta_3$ becomes positive
for all values of $\eta$ and $\tau$, hence the short distance limit of the
function $f(r)$ can be obtained if we expand Eq. (\ref{sol1}) around
the point $r=0.$ The result reads:
\be \label{sdl1}
f(r)\simeq1-\frac{\tilde{C}_{ M}}{\eta}r+\frac{\tau}{\eta}r^2+\frac{5 \eta-\tau^2}{\tilde{C}_{ M}^{1/3} \eta^{5/3}}r^3+O\left(r^4\right)
\ee
The above expansion implies a {\em regular} behavior for $f(r)$ at the axes
origin. It is worth comparing against the corresponding behavior of the solution in the case
of zero cubic terms ($\eta=0$). Note that in the limit $\eta\rightarrow 0$
we cannot use Eq. (\ref{sdl1}), as it blows up for $\eta=0$. However,
in appendix A we have derived the expression in Eq. (\ref{KS}) for $f(r)$ when $\eta=0$ (Kehagias-Sfetsos solution),
hence, if we expand Eq. (\ref{KS}) around the point $r=0$, we obtain
\be \label{sdl2}
f(r)\simeq 1-\sqrt{\frac{\tilde{C}_M}{3 \tau}}\sqrt{r}+\frac{5}{2 \tau}r^2+O\left(r^{7/2}\right)
\ee
This behavior of $f(r)$ has an essential difference with the one in Eq. (\ref{sdl1}).
In particular $f'(r)$, in the case of Eq. (\ref{sdl2}), blows up
as $1/\sqrt{r}$ when
$r$ tends to zero, hence in the absence of the cubic terms the regular point
at $r=0$ turns into a {\em singularity.}

\subsubsection{Zero quadratic terms: $\tau=0$}\label{417}

If we set $\tau= 0$ in equations (\ref{sol1}), (\ref{sval}), (\ref{tval}), we obtain the simplified expression
\be
Z(r)=\left[
\sqrt{\left(\frac{5}{\eta}\right)^3+\left(\frac{\tilde{C}_M}{2 \eta r^3}\right)^2}-\frac{\tilde{C}_M}{2 \eta r^3}\right]^{1/3}-\left[
\sqrt{\left(\frac{5}{\eta}\right)^3+\left(\frac{\tilde{C}_M}{2 \eta r^3}\right)^2}+\frac{\tilde{C}_M}{2 \eta r^3}\right]^{1/3}
\ee
which corresponds to zero quadratic terms in curvature. The above equation
can be written in a more compact form if we set
\be
\tilde{C}_M=30 M, ~~\tilde{\omega}=\frac{5}{\eta}.
\ee
The expression for $f(r)$ reads:
\be
f(r)=1+r \left[\left(\sqrt{\tilde{\omega}^3 r^6+(3\tilde{\omega}M )^2}-3  \tilde{\omega} M\right)^{1/3}-\left(\sqrt{\tilde{\omega}^3 r^6+(3\tilde{\omega} M )^2}+3\tilde{\omega} M \right)^{1/3}\right]
\ee
which has the following large distance asymptotic behavior
\be
f(r)\simeq 1-\frac{2 M}{r}+\frac{8M^3}{3\tilde{\omega} r^8}+O\left(\frac{1}{r^{11}}\right)
\ee
It is worth comparing against the large distance asymptotic behavior of
Kehagias-Sfetsos solution, Eq. (\ref{KS1}) in appendix A , which
is given by the following equation:
\be
f(r)\simeq 1-\frac{2 M}{r}+\frac{2M^2}{\omega r^4}+O\left(\frac{1}{r^{7}}\right)
\ee
Notice the difference in the next-to-leading terms of the above expansions.

\subsubsection{Numerical results for the first case}

\begin{figure}[h]
\begin{center}
\includegraphics[width=0.8 \textwidth, angle=0]{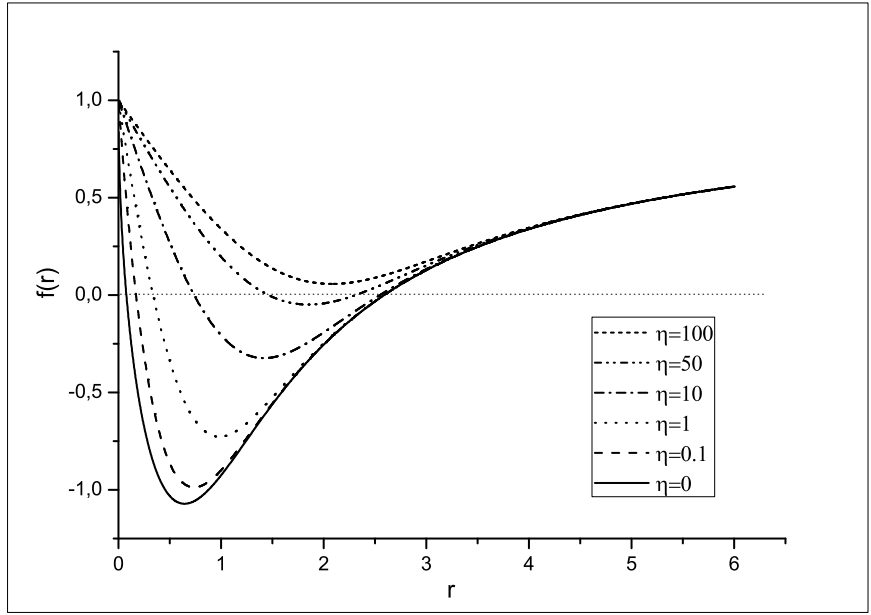}
\end{center}
\caption{\small $f(r)$ versus $r$, for $\tau=1$,
$\eta=0,0.1,1,10,50,100$ and  $\tilde{C}_M=40$
($M=4/3$).}\label{1}
\end{figure}

In this section we study graphically the effects of the cubic
terms, on the short distance behavior of the class of black hole
solutions described in subsections 4.1.1 and 4.1.2. Note, that
$\eta$ measures the effect of cubic in curvature terms, while
$\tau$ corresponds to the effect of second order in curvature
terms.

In Fig. \ref{1} we have plotted the function $f(r)$ for a fixed
value of parameter $\tau=1$, and several values of $\eta:$
$\eta=0,0.1,1,10,50,100$. Notice that $\eta=0$ is the
Kehagias-Sfetsos solution, $\eta=0.1$ corresponds to the solutions
described in subsection 4.1.2, while the remaining values of
$\eta$ correspond to the solutions in subsection 4.1.1. We have
chosen $\tilde{C}_M=40 \Rightarrow M=\frac{\tilde{C}_M}{30} =
4/3,$ which is positive, so the solutions in subsection 4.1.3 are
not shown in this figure. We see that for small values of
$\eta=0.1,1,10$, there are deviations from the $\eta=0$ solution
\footnote{For $\eta=0$ the solution is given by equation
(\ref{KS}) in appendix A.} only inside the horizon. However, for
larger values of $\eta,$ such as $\eta=50$ or $\eta=100,$ when the
effect of cubic in curvature terms becomes stronger, we observe
important differences even outside the Schwarzschild horizon. Note
that in general our solutions have two horizons as shown in Fig.
\ref{1}, but for the large value $\eta=100$ these two horizons
disappear. In addition, as we show in the previous section, for
$\eta \neq 0$, the axes origin is not a naked singularity but a
regular point.

\subsection{Second case: $\varrho=0$, $\eta=0$ and $\tau=0$ } \label{sec42}

If we assume that  $\rho=\eta=\tau=0$ we have $R=0$, hence Eq.
(\ref{Zeq}) can be put into the form
\be
P(Z)\left(r \frac{d Z}{dr}\right)^2+ 4 M(Z) \left(r \frac{d
Z}{dr}\right)+ 4 Q(Z)=0 \label{Zeqsecond} \ee where $P(Z)$,$Q(Z)$
and $M(Z)$ are given by the equations
\be
P(Z)=-\zeta Z+\varphi,~~M(Z)=-5,~~Q(Z)=-15 Z \ee If we solve
(\ref{Zeqsecond}), we obtain \be \label{eqcase2} r
\frac{dZ}{dr}=\frac{\left(10\pm2\sqrt{25+15\varphi Z-15 \zeta
Z^2}\right)}{\varphi-\zeta Z} \ee The above equation has the
correct asymptotic behavior, $H(Z)\simeq -3Z$ for $Z\rightarrow
0$, for the negative sign, so we will only analyze only this
branch. If  we integrate, assuming that $\zeta<0$ (that is,
setting $\zeta=-|\zeta|$), we obtain:
\begin{eqnarray} \label{sol2}
&&\left| \sqrt{1+\frac{3}{5}\varphi Z+\frac{3}{5}|\zeta| Z^2}+\sqrt{\frac{3}{5}|\zeta|}Z+\sqrt{\frac{15}{|\zeta|}}\frac{\varphi}{10 }\right|^{\sqrt{\frac{15}{|\zeta|}}\frac{\varphi}{10 }}\left|\sqrt{1+\frac{3}{5}\varphi Z+\frac{3}{5}|\zeta| Z^2}-\frac{3}{10} \varphi Z-1\right| \times \nonumber \\ &&\times e^{\left(\sqrt{1+\frac{3}{5}\varphi Z+\frac{3}{5}|\zeta| Z^2}-1\right)}=\frac{\tilde{C}^2_M}{r^6}
\end{eqnarray}
If we set
\be
\tilde{C}^2_{M}=M^2
\left|\frac{6|\zeta|}{5}-\frac{9\phi^2}{50}\right|~~ \left|
1+\sqrt{\frac{15}{|\zeta|}}\frac{\varphi}{10
}\right|^{\sqrt{\frac{15}{|\zeta|}}\frac{\varphi}{10 }} \ee we
recover the asymptotic behavior $Z\rightarrow -2M/r^3$ for
$r\rightarrow +\infty$. For the function $\tilde{N}(Z)$, we
obtain: from Eqs. (\ref{g01}), (\ref{gZ1}) and (\ref{gZ2})
\begin{eqnarray} \label{Ncase2}
\tilde{N}^2(Z)&=&\tilde{C}_{N}\frac{e^{\sqrt{1+\frac{3}{5}\varphi Z+\frac{3}{5}|\zeta| Z^2}-1}}{\left(1+\frac{3}{5}\varphi Z+\frac{3}{5}|\zeta| Z^2\right)} \left| \sqrt{1+\frac{3}{5}\varphi Z+\frac{3}{5}|\zeta| Z^2}+\sqrt{\frac{3}{5}|\zeta|}Z+\sqrt{\frac{15}{|\zeta|}}\frac{\varphi}{10 }\right|^{\sqrt{\frac{15}{|\zeta|}}\frac{\varphi}{10 }}\nonumber \\ &&\times\left|\sqrt{1+\frac{3}{5}\varphi Z+\frac{3}{5}|\zeta| Z^2}-\frac{3}{10}\varphi \left(Z+\frac{\varphi}{\zeta}\right)+1\right|
\end{eqnarray}
in which $\tilde{C}_{N}$ is an integration constant equal to
\be
\tilde{C}_{N}=\left|2-\frac{3}{10}\frac{\varphi^2}{|\zeta|}\right|^{-1}\left|
1+\sqrt{\frac{15}{|\zeta|}}\frac{\varphi}{10
}\right|^{-\sqrt{\frac{15}{|\zeta|}}\frac{\varphi}{10 }} \ee if we
demand $\tilde{N}\rightarrow 1$ for $Z\rightarrow 0$,  as it is
required in order to achieve $N^2(r)\simeq f(r)$ for large
distances.  Finally, we note that $N(r)^2$ blows up like $1/r^6$
as $r$ tends to zero (or $Z\rightarrow -\infty$), which implies a
space time singularity at the axes origin.

\subsubsection{Solution for zero quadratic terms}\label{421}

If we set $\varphi=0$ in Eq. (\ref{sol2}) we obtain \be
\label{sol2s20} \left(\sqrt{1+\frac{3}{5}|\zeta|
Z^2}-1\right)~~e^{\left(\sqrt{1+\frac{3}{5}|\zeta|
Z^2}-1\right)}=\frac{\tilde{C}^2_M}{r^6} \ee which can be solved
analytically. The solution of the above equation can be expressed
via the Lambert function $W_L(x)$, which is defined as the real
solution of the equation $W_L(x) e^{W_L(x)}=x$, hence \be
\label{cubic1} \sqrt{1+\frac{3}{5}|\zeta| Z^2}-1=
W_L\left(\frac{\tilde{C}^2_M}{r^6}\right) \ee Simple calculations
yield:
\be
Z=\pm\left(\frac{5}{3|\zeta|}\right)^{1/2}
\sqrt{W_L^2\left(\frac{\tilde{C}^2_M}{r^6}\right)+ 2
W_L\left(\frac{\tilde{C}^2_M}{r^6}\right)} \ee and finally
\be
f(r)=1+r^2 Z=1-r^2 \left(\frac{5 }{3|\zeta|}\right)^{1/2}
\sqrt{W_L^2\left(\frac{\tilde{C}^2_M}{r^6}\right)+2
W_L\left(\frac{\tilde{C}^2_M}{r^6}\right)} \ee Note that we have
chosen the solution with the negative sign in order to achieve
positive gravitational mass. For the large and short distance
limits of $f(r)$ we can use the properties:
\begin{eqnarray}
&& x\rightarrow 0 \Rightarrow W_{L}(x)\simeq x \\
&& x\rightarrow +\infty \Rightarrow W_{L}(x)\simeq \ln(x)
\end{eqnarray}
so we obtain for $r\rightarrow +\infty:$
\begin{equation}
f(r)\simeq 1-\left(\frac{10 }{3|\zeta|}\right)^{1/2}  \frac{\tilde{C}_{M}}{r}+O\left(\frac{1}{r^7}\right)
\end{equation}
hence for the mass of the black hole we get
\be
M=\left(\frac{5 }{6|\zeta|}\right)^{1/2} \tilde{C}_{M} \ee which
is assumed to be positive. Now, for $r \rightarrow 0$ we find
that:
\be
f(r)\simeq 1-r^2 \left(\frac{5 }{3|\zeta|}\right)^{1/2}
\sqrt{\ln^2\left(\frac{\tilde{C}^2_M}{r^6}\right)+2
\ln\left(\frac{\tilde{C}^2_M}{r^6}\right)} \ee The above
expression yields a second derivative which blows up at the point
$r=0$, hence this point corresponds to a spacetime singularity.
For the function $N(r)$, if we set $\varphi=0$ in Eq.
(\ref{Ncase2}) we obtain
\be
\tilde{N}^2(Z)=\frac{e^{\sqrt{1+\frac{3}{5}|\zeta|
Z^2}-1}}{2\left(1+\frac{3}{5}|\zeta|
Z^2\right)}\left(1+\sqrt{1+\frac{3}{5}|\zeta| Z^2}\right) \ee
Taking into account Eq. (\ref{cubic1}) and the definition of
Lambert function, we obtain the following closed form \be
\label{Ncubic} N(r)^2=f(r) \tilde{N}(Z(r))^2=\frac{\bar{C}_{M}^2
\left(2+W_{L}\left(\frac{\tilde{C}^2_M}{r^6}\right)\right) f(r)
}{2 r^6 W_{L}\left(\frac{\tilde{C}^2_M}{r^6}\right)\left(1+
W_{L}\left(\frac{\tilde{C}^2_M}{r^6}\right) \right)^2} \ee In the
large $r$ regime we find from the above equation that: \be
\label{asymN} N(r)^2=f(r)~\left(1-\frac{3}{2}\frac{\bar{C}_{M}^2
}{ r^6}+O\left(\frac{\bar{C}_{M}^4}{r^{12}}\right)\right) \ee
hence in the large distance limit we recover the standard
asymptotic behavior $N(r)^2\simeq f(r)$. In addition, note that
$N(r)^2$ blows up when r tends to zero ($Z\rightarrow -\infty$)
like $1/r^6$, as we see in Eq. (\ref{Ncubic}) above.

\subsubsection{Numerical results for the second case}

\begin{figure}[h]
\begin{center}
\includegraphics[width=0.8 \textwidth, angle=0]{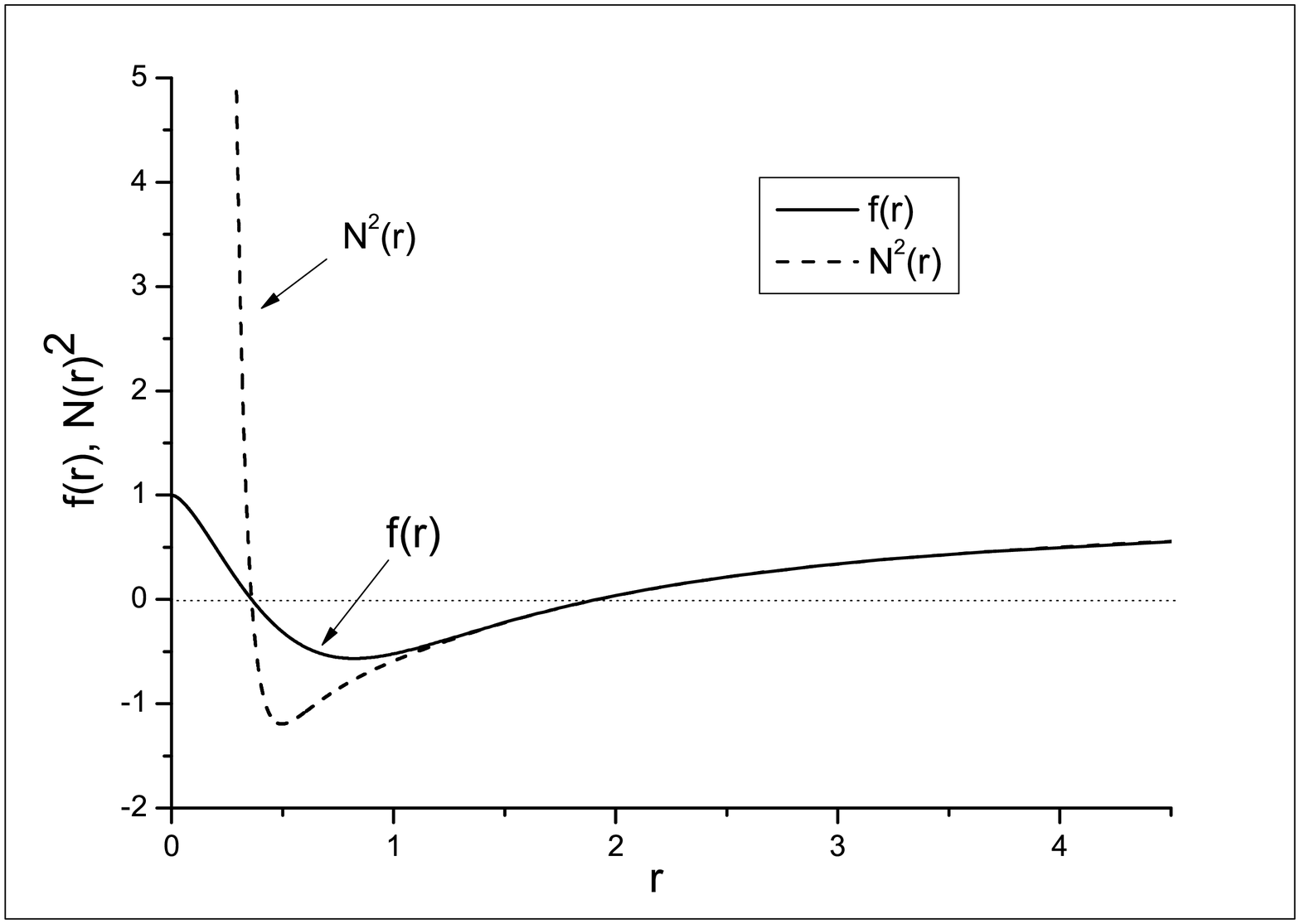}
\end{center}
\caption{\small $f(r)$ and $N(r)^2$ versus $r$, for $\phi=-1$, $|\zeta|=0.5$ and  $M=1$.}\label{2}
\end{figure}

In this section we have used the analytical expressions of Sec.
4.2 to illustrate graphically the black hole solutions. As we see
in Fig. \ref{2} the black hole solution has the familiar form with
two horizons, but in contrast with the solution presented in Sec.
4.1, the function $N(r)^2$ differs significantly from $f(r)$
inside the outer horizon and blows up for $r=0$, which implies a
spacetime singularity. Note that the solution for the first case,
presented in Sec. 4.1, are regular in the axes origin.

\begin{figure}[h]
\begin{center}
\includegraphics[width=0.8 \textwidth, angle=0]{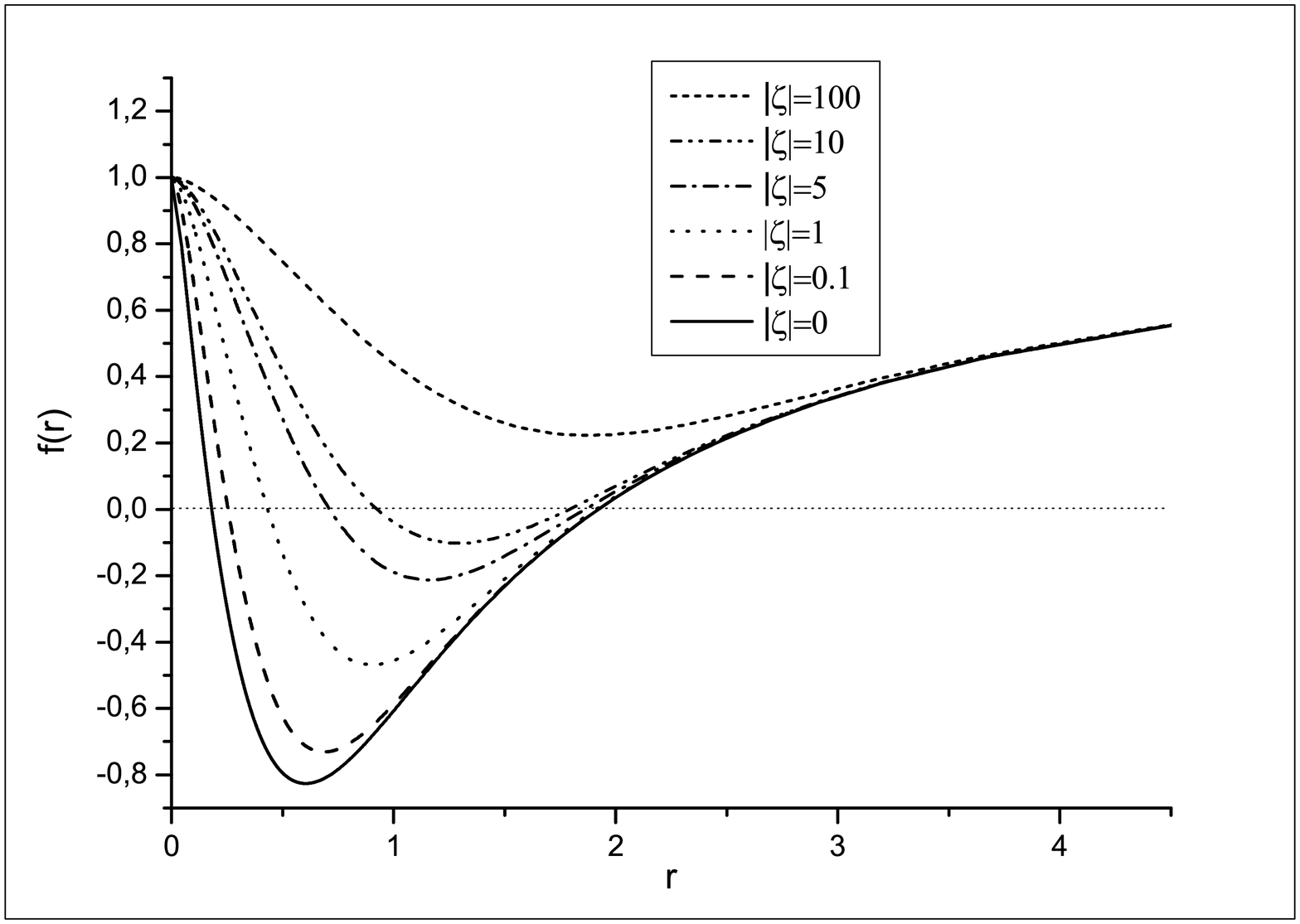}
\end{center}
\caption{\small $f(r)$ versus $r$, for $\phi=-1$,
$|\zeta|=0,0.1,1,5,10,100$ and  $M=1$.}\label{3}
\end{figure}

In Fig. \ref{3} we have plotted the function $f(r)$ for a fixed
value of parameter $\phi=-1$, and several values of
$|\zeta|=0,0.1,1,5,10,100$, for $M=1.$ We observe that when the
effect of cubic terms becomes stronger (the parameter $|\zeta|$
becomes larger), the two horizons coalesce into one and finally
disappear, hence for large $|\zeta|$ our solution represents a
naked singularity.

\section{Conclusions}

We considered four-dimensional non-projectable Horava-Lifshitz
type gravity, without detailed balance, for an action with all
possible terms quadratic and cubic in the curvature. Black hole
solutions in the case of quadratic terms have been
studied previously in the literature, however in this work we examine for
the first time the cubic terms. Although in the general case it is
not possible to find analytic solutions, for special choices of
the free parameters of the model, we obtained two new black hole
solutions which exhibit the standard Schwarzschild asymptotic
behavior in the large distance limit.

It is worth mentioning that the two solutions we have obtained,
reduce to known solutions of Horava-Lifshitz gravity when the
cubic terms are turned off. In particular, the solution of Sec.
\ref{sec41} corresponds to Kehagias-Sfetsos (KS) solution (see
Appendix A) and the solution in Sec. \ref{sec42} corresponds to
Kiritsis-Kofinas (KK) solution (see Appendix B). In Figs. \ref{1}
and \ref{3} the effects of cubic terms are illustrated, so the
reader can compare with KS and KK black hole solutions in a
straightforward fashion.

In addition, we stress that the cubic black hole solutions we
obtained in Sec. \ref{sec41}  have a regular center, in contrast
with the KK solution, in which the center is singular. On the
other hand, in the case of the second solution described in
Sec.\ref{sec42}, the cubic terms are not enough to cure the
singularity (see Fig. \ref{2}). Note, that a similar singularity
is present in the KK black hole solution in which only quadratic
terms in curvature present.

Finally, in Sec. \ref{417} and \ref{421} we present the solutions
when quadratic terms are turned off. In this case the analytical
formulae for the solutions are considerably simplified. In
addition note, that the short and large distance behavior of these
purely cubic solutions differs from the corresponding behavior of
KS and KK black holes. Further investigation on these black hole
solutions may be a topic in a future publication.

\section{Acknowledgements}
We would like to thank the anonymous referee for his valuable comments on the manuscript.

\appendix

\section{Appendix: Zero cubic terms ($\varrho=0$, $\eta=0$ and $\zeta=0$)}\label{appA}

Of special interest is the case where one has ho cubic terms at
all. These solutions has already been studied by A.Kehagias and K.Sfetsos.
in Ref. \cite{Kehagias:2009is} and by E. Kiritsis and G. Kofinas in \cite{Kiritsis:2009rx}.

\subsection{Kehagias-Sfetsos solution: $\varphi=0$ and $\tau\neq 0$}\label{appA1}

If we set $\varphi=0$ in (\ref{Zeq0}) we obtain:
\be
3 \tau Z^2-15 Z-\frac{\tilde{C}_{M}}{r^3}=0 \ee with the two
solutions
\be
Z=\frac{5}{2 \tau}\left(1\pm\sqrt{1+\frac{4 \tau \tilde{C}_{M}}{75
r^3}}\right). \ee For the function $f=1+r^2 Z$ we get: \be
\label{KS} f(r)=1+r^2\frac{5}{2 \tau}\left(1-\sqrt{1+\frac{4 \tau
\tilde{C}_{M}}{75 r^3}}\right), \ee where we have kept the
negative sign to have the standard $1-2M/r$ behavior
asymptotically. Also, note that $\tau \tilde{C}_{M}$ should be
positive, otherwise the solution has no short-distance limit.

If $r\rightarrow +\infty$ we find the following asymptotic formula
\be
f(r)\simeq1-\frac{ \tilde{C}_{M}}{15 r}, \ee hence for the mass of
the black hole we obtain \be \label{mass}
M=\frac{\tilde{C}_{M}}{30 }. \ee This relation, along with the
notation
\be
\omega=\frac{5}{2 \tau}, \ee transforms (\ref{KS}) into: \be
\label{KS1} f(r)=1+\omega r^2-\sqrt{r\left(\omega^2 r^3+4 \omega
M\right) }, \ee which is the standard form as presented in
\cite{Kehagias:2009is}. For the function $N(r)$ we know that
\begin{equation}
N(r)^2=f(r)
\end{equation}
Note that the recovering of the SK solution for $\varphi=0$ is expected, as
the action we consider in Sec. 2 when the cubic terms are absent, is identical with the action for the detailed balance
condition for $\lambda=1$. In the following Appendix \ref{appB} we give a detail analysis on this topic.

\subsection{Kiritsis-Kofinas solution: $\tau=0$ and $\varphi\neq 0$ }\label{appA2}

If we set $\zeta=0$ in Eq. (\ref{eqcase2}) we find
\be
r \frac{dZ}{dr}=\frac{1}{\varphi}\left(10\pm2\sqrt{25+15\varphi
Z}\right) \ee If we integrate the above equation, in the case of
the negative sign, which corresponds to asymptotically flat
space-time, we obtain: \be \label{quadr2}
\left(\sqrt{1+\frac{3}{5}\varphi
Z}-1\right)e^{\sqrt{1+\frac{3}{5}\varphi
Z}-1}=\frac{\bar{C}_{M}}{r^3} \ee where $\bar{C}_{M}$ is a
positive constant of integration, related to the mass $M$ of the
black hole, as we will see. From the definition of Lambert
functions, as the real solution of the equation $W_{L}(x)
e^{W_{L}(x)}=x$, we find
\be
\sqrt{1+\frac{3}{5}\varphi
Z}-1=W_{L}\left(\frac{\bar{C}_{M}}{r^3}\right), \ee which yields:
\be
Z=\frac{5}{3 \varphi}
\left(W_{L}^2\left(\frac{\bar{C}_{M}}{r^3}\right)+2
W_{L}\left(\frac{\bar{C}_{M}}{r^3}\right)\right), \ee while for
$f(r)$ we get
\be
f(r)=1+\frac{5 r^2}{3 \varphi}
\left(W_{L}^2\left(\frac{\bar{C}_{M}}{r^3}\right)^2+2
W_{L}\left(\frac{\bar{C}_{M}}{r^3}\right)\right) \ee In order to
obtain the large and short distance limits of $f(r)$ we can use
that:
\begin{eqnarray}
&& x\rightarrow 0 \Rightarrow W_{L}(x)\simeq x \\ && x\rightarrow
+\infty \Rightarrow W_{L}(x)\simeq \ln(x).
\end{eqnarray}
For $r\rightarrow +\infty$ we obtain
\begin{equation}
f(r)\simeq 1+\frac{10}{3 \varphi  } \frac{\bar{C}_{M}}{r}+O\left(\frac{1}{r^4}\right)
\end{equation}
and the mass of the black hole reads:
\be
M=-\frac{10\bar{C}_{M} }{6 \varphi} \ee which is assumed to be
positive, or $\varphi <0$. For $r\rightarrow 0$ we find that:
\be
f(r)\simeq 1+\frac{5 r^2}{3 \varphi}
\left(\ln^2\left(\frac{\bar{C}_{M}}{r^3}\right)^2+2
\ln\left(\frac{\bar{C}_{M}}{r^3}\right)\right) \ee The above
expression yields a second derivative which blows up at the point
$r=0$, hence this point corresponds to a spacetime singularity.
For the function $N(r)$, if we use Eqs. (\ref{g01}), (\ref{gZ1})
and (\ref{gZ2}) we obtain
\be
\tilde{N}(Z)=\frac{e^{\sqrt{1+\frac{3}{5}\varphi
Z}-1}}{\sqrt{1+\frac{3}{5}\varphi Z}} \ee If we take into account
Eq. (\ref{quadr2}) and the definition of the Lambert function, we
obtain the closed form \be \label{Nquadr} N(r)^2=f(r)
\tilde{N}(Z(r))^2=\frac{\bar{C}_{M}^2 f(r) }{r^6
\left(W_{L}^2\left(\frac{\bar{C}_{M}}{r^3}\right)+
W_{L}\left(\frac{\bar{C}_{M}}{r^3}\right) \right)^2} \ee In the
large $r$ regime we find, from the above equation, that: \be
\label{asymN} N(r)^2=f(r)~\left(1-2\frac{\bar{C}_{M} }{
r^3}+O\left(\frac{\bar{C}_{M}^2}{r^{6}}\right)\right) \ee hence in
the large distance limit we recover the standard asymptotic
behavior $N(r)^2\simeq f(r)$. In addition, note that $N(r)^2$
blows up when r tends to zero like $1/r^6$, as we see in Eq.
(\ref{Nquadr}) above.

\section{Appendix B: ~Comparing with the detailed balance action}\label{appB}

In this appendix we present briefly the "detailed balance condition", which a method for the construction of the potential term in the Lagrangian of HL gravity, see Sec. 2.
It is inspired by nonequilibrium statistical physics, and it was proposed in Ref. \cite{Horava:2008ih}.
The main advantage of this approach is the restriction of the large number of arbitrary couplings that appear in the bare action of the model.
In addition, if the action of the model has been constructed by using the detailed balance, from an original $d$ dimensional action, then the corresponding $d+1$ field theory model has improved renormalization properties.
These properties have been inherited from the original $d$ dimensional action, see for example Ref. \cite{Anagnostopoulos:2010gw} and references there in.

In the case of the detailed balance condition the potential term, of the d+1 dimensional action, is derived from a variation principle, or equivalently by taking the square of the equation of motion of an original Euclidian d dimensional action. Particularly, in the case HL gravity model the action reads:
\begin{equation} \label{app1}
S_{V}=\frac{\kappa^2}{2}\int dt d^d x \sqrt{|g|} N {\cal L}_{V}, ~~~d=3
\end{equation}
in which $\kappa^2$ is a coupling constant, where
\begin{equation}\label{app2}
{\cal L}_{V}= E^{ij} \; {\cal G}_{ij;kl} \: E^{kl}, ~~~ E^{ij}=\frac{1}{\sqrt{|g|}}  \frac{\delta W[g_{kl}]}{\delta g_{ij}}
\end{equation}
and  ${\cal G}_{ij;kl}$ is the inverse generalized De Witt metric which is defined as
\begin{equation}\label{app3}
{\cal G}^{ij;kl}=\frac{1}{2}\left(g^{ik}g^{jl}+g^{il}g^{jl}\right)-\lambda g^{ij}g^{kl}
\end{equation}
where $\lambda$ is a new dynamical coupling, underlie to quantum corrections.
The standard choice for the three dimensional (d=3) action $W[g_{kl}]$ is
\begin{equation} \label{app4}
W[g_{kl}]=W_{C}[g_{kl}]+W_{EH}[g_{kl}], ~  W_{C}=\frac{1}{w^2}\int d^3x \; \omega_{3}(\Gamma), ~  W_{EH}=\mu \int d^3x \sqrt{|g|}(R-2\Lambda_{W})
\end{equation}
in which
\begin{equation}\label{app5}
\omega_{3}(\Gamma)=\varepsilon^{ijk}\left(\Gamma_{il}^{m}\partial_{j}\Gamma_{km}^{l}+\frac{2}{3}\Gamma_{il}^{n}\Gamma_{jm}^{l}\Gamma_{kn}^{m}\right)
\end{equation}
and $w^2$ is a dimensionless coupling, while $\mu$ and $\Lambda_{W}$ are couplings with dimensions $[\mu]=1$ and $[\Lambda_{W}]=2$.
Note, that the Cotton tensor, which is defined as
\begin{equation}\label{app6}
C_{ij}=\varepsilon^{ikl} \nabla_{k}\left(R_{l}^{j}-\frac{1}{4}R \delta_{l}^{j}\right),
\end{equation}
can be obtain from the action $W_{C}[g_{kl}]$, more specifically
\begin{equation}\label{app7}
\frac{1}{\sqrt{|g|}} \frac{\delta W_C[g_{kl}]}{\delta g_{ij}}=\frac{1}{w^2}C_{ij}
\end{equation}
For the tensor $E_{ij}$ in Eq. (\ref{app2}), we find
\begin{equation}
E_{ij}=\frac{1}{w^2}C_{ij}-\frac{\mu}{2}\left(R^{ij}-\frac{1}{2}R g^{ij}+\Lambda_{W}g^{ij}\right)
\end{equation}
Now, for the potential part of the Lagrangian (see Eqs. (\ref{app1}), (\ref{app2}) and (\ref{app3})) we obtain
\begin{equation}
{\cal L}_{V}=-\frac{1 }{ w^4} C_{ij}C^{ij}+\frac{ \mu}{ w^2} \varepsilon^{ijk}R_{il} \nabla_{j}R_{k}^{l}-\frac{\mu^2}{4}R_{ij}R^{ij}+\frac{\mu^2}{4(1-3\lambda)}\left(\frac{1-4\lambda}{4}R^2+\Lambda_{W}R-3 \Lambda_{W}^2\right)
\end{equation}
As we are looking for spherically symmetric solutions without a cosmological constant (the spacetime should be Minkowsky asymptotically) it is reasonable to set $\Lambda_{W}=0$ in the above action. Moreover in
order to recover general relativity in the IR limit we have to add in the action a relevant term proportional to $\mu^4 R$. This term violates the detail balanced in general,
but this term becomes unimportant in the UV where the theory still satisfies the detailed balance condition. Finally, we take
\begin{eqnarray} \label{app10}
{\cal L}_{V}=-\frac{1 }{w^4} C_{ij}C^{ij}+\frac{ \mu}{w^2} \varepsilon^{ijk}R_{il} \nabla_{j}R_{k}^{l}-\frac{\mu^2}{4}R_{ij}R^{ij}+\frac{\mu^2(1-4\lambda)}{16(1-3\lambda)}R^2+\frac{\kappa^2\mu^4}{2} R
\end{eqnarray}
The corresponding full action with the kinetic term can be constructed as it is presented in Sec. 2.
Comparing with Eq. (\ref{IR}) in Sec. 2, which is valid in the IR limit, we take:
\be
\lambda=1, ~~~~c^2=\frac{\kappa^2\mu^4}{2},~~~~ G=\frac{\kappa^2}{32 \pi c}
\ee
We stress that in the case of spherically symmetric solutions, more specifically for a metric of the form of Eq. (\ref{metric}), or
\be
\label{app11}
 ds^2=-N(r)^2 dt^2+\frac{dr^2}{f(r)} +r^2 d\Omega^2
\ee
the first two terms in Eq. (\ref{app10}), which determine the UV behavior of the model, vanish identically, or we have
\be
\label{app12}
-\frac{1 }{ w^4} C_{ij}C^{ij}=0, ~~~\frac{ \mu}{ w^2} \varepsilon^{ijk}R_{il} \nabla_{j}R_{k}^{l}=0
\ee
Now, we can compare straightforwardly, the Lagrangian of Eq. (\ref{app10}) without the terms which are proportional to  $C_{ij}C^{ij}$ and $\varepsilon^{ijk}R_{il} \nabla_{j}R_{k}^{l}$,
with the Lagrangian of Sec. 2 that reads:
\be
{\cal L}_V=\eta_{1a} R+\eta_{2a} R^2+\eta_{2b} R^{ij}R_{ij}
\ee
when the cubic and derivative terms are absent (${\cal L}_{R^3}=0$ and  ${\cal L}_{\Delta R^2}=0$, see Sec. 2), then we find
\be
\eta_{2b}=-\frac{\mu^2}{4},~~\eta_{2a}=\frac{\mu^2(1-4\lambda)}{16(1-3\lambda)}
\ee
For $\lambda=1$, we obtain that $$\frac{\eta_{2a}}{\eta_{2b}}=-\frac{8}{3},$$ or equivalently the coupling $\varphi=40 \eta_{2a}+15 \eta_{2b}$ is equal to zero.
We conclude that the condition $\varphi=0$ corresponds \footnote{This result is valid only for metrics with spherical symmetry.} to the detail balance condition with $\lambda=1$, therefore the solution we found
in Appendix \ref{appA1}, in the case of zero cubic terms, is expected to be identical with SK solution of Ref. \cite{Kehagias:2009is}.

\end{document}